# Modeling and Analysis of Integrated Proactive Defense Mechanisms for Internet-of-Things

Mengmeng Ge[1], Jin-Hee Cho[2], Bilal Ishfaq[3], and Dong Seong Kim[4]

[1]Deakin University, Geelong, VIC, Australia

[2]Virginia Tech, Falls Church, VA, USA

[3]University of Canterbury, Christchurch, New Zealand

[4]University of Queensland, Brisbane, QLD, Australia

**Abstract –** As a solution to protect and defend a system against inside attacks, many intrusion detection systems (IDSs) have been developed to identify and react to them for protecting a system. However, the core idea of an IDS is a reactive mechanism in nature even though it detects intrusions which have already been in the system. Hence, the reactive mechanisms would be way behind and not effective for the actions taken by agile and smart attackers. Due to the inherent limitation of an IDS with the reactive nature, intrusion prevention systems (IPSs) have been developed to thwart potential attackers and/or mitigate the impact of the intrusions before they penetrate into the system. In this chapter, we introduce an integrated defense mechanism to achieve intrusion prevention in a software-defined Internet-of-Things (IoT) network by leveraging the technologies of cyberdeception (i.e., a decoy system) and moving target defense, namely MTD (i.e., network topology shuffling). In addition, we validate their effectiveness and efficiency based on the devised graphical security model (GSM)-based evaluation framework. To develop an adaptive, proactive intrusion prevention mechanism, we employed fitness functions based on the genetic algorithm in order to identify an optimal network topology where a network topology can be shuffled based on the detected level of the system vulnerability. Our simulation results show that GA-based shuffling schemes outperform random shuffling schemes in terms of the number of attack paths toward decoy targets. In addition, we observe that there exists a tradeoff between the system lifetime (i.e., mean time to security failure) and the defense cost introduced by the proposed MTD technique for fixed and adaptive shuffling schemes. That is, a fixed GA-based shuffling can achieve higher MTTSF with more cost while an adaptive GA-based shuffling obtains less MTTSF with less cost.

## 10.1 Introduction

Internet-of-Things (IoT) has received significant attention due to their enormous advantages. Advances in IoT technologies can be easily leveraged to maximize effective service provisions to users. However, due to the high heterogeneity and the constrained resources of composed entities, and its large-scale networks, we face the following challenges [23]: (1) distributed technologies for communications, data filtering, processing, and dissemination with an enormous amount of various forms of data (e.g., text, voice, haptics, image, video) in large-scale networks with heterogeneous entities (i.e., devices, humans); (2) severely restricted resources in battery, computation, communication (e.g., bandwidth), and storage, causing significant challenges in resource allocation and data processing capabilities; (3) highly adversarial environments, introducing compromised, deceptive entities and data, which may result in detrimental impacts on the capabilities of critical mission-related decision making; and (4) highly dynamic interactions between individual entities, data, and environmental factors (e.g., network topology or resource availability), where each factor is highly dynamic in time/space. Due to these characteristics of IoT environments, highly secure, lightweight defense mechanisms are in need to protect and defend a system (or network) against potential attacks.

As a solution to protect and defend a system against inside attacks, many intrusion detection systems (IDSs) have been developed to identify and react to the attacks. However, the core idea of an IDS is reactive in nature and even though it detects intrusions which have already been in the system. Hence, this reactive mechanism would be way behind and not effective for the actions by agile and smart attackers. Due to the inherent limitation of an IDS with these reactive nature, intrusion prevention systems (IPSs) have been developed to thwart potential attackers and/or mitigate the impact of the intrusions before they penetrate into the system [5]. In this work, we are interested in developing an integrated intrusion prevention mechanism based on the technologies of cyberdeception (i.e., a decoy system) and moving target defense, namely MTD (i.e., network topology shuffling), and evaluating their effectiveness and efficiency by a graphical security model (GSM)-based evaluation framework.

### 10.1.1 Research Goal & Contributions

This work aims to propose an integrated proactive defense based on intrusion preventive mechanisms, such as cyberdeception and MTD techniques, to minimize the impact of potential attackers trying to penetrate into IoT systems via multiple entries. We make the following ***key contributions*** in this book chapter:

- We developed an *integrated proactive defense system* by proposing an adaptive MTD technique by shuffling a network topology where a network consists of both decoy nodes and real nodes. As decoy nodes are the part of a decoy system, which is a common cyberdeception technology, this work integrates an MTD technique with cyberdeception to propose an adaptive, proactive intrusion defense mechanism that maximizes hurdles and/or complexity for attackers to launch their attacks while minimizing the defense cost to execute MTD operations. The key goal of the proposed network topology shuffling-based MTD (NTS-MTD) with decoy nodes is to generate a network topology that can maximize disadvantages against the attackers. Little work has integrated both cyberdeception and MTD techniques particularly in terms of network topology shuffling in software-defined networking (SDN)-based IoT environments.

- We took a metaheuristic approach based on a genetic algorithm (GA) by devising fitness functions that can achieve the objective of our proposed proactive defense mechanism in terms of minimizing defense cost as well as security vulnerability in attack paths. For example, given a set of network topologies that are uniformly selected at random, we use the GA and devise fitness functions that identify an optimal network topology to minimize security vulnerability and defense cost (e.g., network shuffling cost). In particular, the devised fitness functions estimate the utility of triggering NTS-MTD operation based on a generated network topology in an adaptive way, instead of triggering the MTD operation based on a fixed time interval. This allows our proposed integrated defense mechanism to provide a *secure, affordable defense service*.

- We considered a *SDN-based IoT* as our network environment. The merits of SDN technology in terms of programmability and controllability are highly leveraged to design the proposed proactive defense mechanism equipped with MTD and cyberdeception techniques and examine its performance in terms of security and performance metrics, including the number of attack paths towards decoy targets, mean time to security failure (i.e., MTTSF or system lifetime), and defense cost.

- We adopted a graphical security model, namely GSM [14], [12], to evaluate the proposed deception and MTD techniques. The GSM offers design solutions to consider attack graphs (AGs) and/or attack trees (ATs) which can provide efficient methods to calculate the potential security (or vulnerability) levels of attack paths. In order to deal with large-scale networks, we will develop a Hierarchical Attack Representation Model (HARM) [12], [14] as a GSM model. The HARM allows us to evaluate the proposed proactive mechanisms including both the cyberdeception and MTD techniques.

**10.1.2 Structure of This Chapter**

The rest of this chapter is organized as follows. Section 10.2 provides the brief overview of the related work in terms of MTD and cyberdeception techniques for IoT environments, security models and metrics, and SDN technology and its use for IoT environments. Section 10.3 gives the overview of the system model, including models of describing a targeted network environment, node characteristics, attack behaviors, and defense mechanisms in place and security failure conditions. Section 10.4 describes the design features of the proposed integrated proactive defense mechanism in terms of NTS-MTD, optimization techniques for NTS-MTD based on GA, and GSM for NTS-MTD in IoT environments. Section 10.5 shows the experimental results and discusses the overall trends of the results observed. Section 10.6 concludes this chapter and suggests future work directions.

## 10.2 Related Work

In this section, we provide the overview of the state-of-the-art approaches with the topics as the basis of the proposed work, including: (1) MTD and cyberdeception techniques for IoT; (2) security models and metrics; and (3) SDN technology for IoT.

### 10.2.1 MTD and Cyberdeception Techniques for IoT

**Moving target defense (MTD)** is an emerging technique aiming at constantly changing the attack surface of the networks to increase the attack complexity [14]. Based on [14], MTD approaches are classified into three categories: shuffling, diversity and redundancy. Shuffling changes the network configurations (e.g., IP address randomization, device migration, or topology reconfiguration). Diversity employs a variety of different implementations for the same functionalities (e.g., choosing various operating systems for the web server). Redundancy provides replications of the devices in the network to increase network reliability in the presence of attacks.

MTD approaches have been proposed to protect resource-constrained IoT devices. In our prior work [12], we examined the performance of address space layout randomization (ASLR) for wireless sensor nodes as an MTD technique and evaluated its effectiveness using the proposed HARM in one of the use cases. Casola et al. [4] proposed an MTD framework to reconfigure the IoT devices by switching among different cryptosystems and firmwares, and then evaluated the framework via a case study of wireless sensor networks. Two metrics, attack probability and attack time, are used to assess the effectiveness of the reconfiguration; but there is still a lack of system-level metrics to quantify the proposed approach. Sherburne et al. [27] proposed a dynamically changing IPv6 address assignment approach over the IoT devices using Low-Powered Wireless Personal Area Networks (LPWPANs) protocol to defend against various network attacks. Zeitz et al. [30] extended [27] and presented a design of fully implementing and testing the MTD approach, which uses the address rotation to obscure the communications among IoT devices. However, both works [27], [30] do not have any experimental validation of the design. Mahmood et al. [19] developed an MTD security framework based on context-aware code partitioning and code diversification for IoT devices to obfuscate the attackers; but it also does not perform any analysis to validate the framework.

**Cyberdeception techniques** are proactive approaches in cyber defense which add an extra layer of defense onto the traditional security solutions (e.g., Intrusion Detection System, or IDS, firewalls, or endpoint anti-virus software). Once attackers are inside a network, they start probing to acquire information to determine the potential valuable assets and then move laterally in the network to launch attacks based on the information they gather during the probes. Cyber deception aims at luring the attackers into the decoy systems within the network and interacting with the attackers to monitor and analyze attackers' behaviors. Honeypot is one of the commonly used technologies in cyberdeception. It is created as a fake asset and deployed around the valuable assets to divert attackers. However, the management complexity and scalability issues of the honeypots hinder the wide usage by the enterprises. Modern deception technology uses basic honeypot technology along with visualization and automation techniques [20]. It allows distributed deployment and update of decoy systems to achieve adequate coverage but still cost-effective.

The state-of-the-art approaches in this domain have focused on developing and deploying honeypot technology for IoT. La et al. [16] introduced a game theoretic method to model the interaction between an attacker who can deceive the defender with suspicious or seemingly normal traffic and a defender in honeypot-enabled IoT networks. Anirudh et al. [2] used honeypots for online servers to mitigate Distributed Denial of Service (DDoS) attacks launched from IoT devices. Dowling et al. [8] created a ZigBee honeypot to capture attacks and used it to identify the DDoS attacks and bot malware. However, none of the work analyzed the impact of the deception techniques on the system-level security. Besides, little study analyzed the modern deception technology for an IoT system which allows distributed deployment and the update of decoys to achieve adequate coverage and provide cost-effective defense service [22].

In our prior work [5], we looked into an integrated defense system to identify what components of each defense mechanism can provide a best solution for achieving "defense in breadth" considering both enhanced security and defense cost. However, it is purely a model-based analysis based on an abstract model in which a granularity of each defense mechanism is omitted to capture a system-level performance to some extent. In addition, there is no current work on developing an integrated defense system equipped with two proposed defense techniques in an IoT, including deception and MTD techniques. Therefore, this work proposes integrated and proactive defense mechanisms that can effectively and efficiently mitigate the adverse effect of attackers before the attackers penetrate into a target IoT system.

### 10.2.2 Security Models and Metrics

**Graphical security models** (attack graphs (AGs) [28], attack trees (ATs) [25]) have been widely employed for security analysis in various types of networks and combined with security metrics to evaluate different defense mechanisms. In the graph-based attack models, an AG shows all possible sequences of the attackers' actions that eventually reach the target. With the increasing size of the network, calculation of a complete AG has exponential complexity, thus causing a scalability issue. In the tree-based attack models, an AT is a tree with nodes representing the attacks and the root representing the goal of attacks. It systematically presents potential attacks in the network. However, an AT also has the scalability issue when the size of the network increases. In order to address the above issues, the two-layered HARM [14] was introduced, which combines various graphical security models onto different layers. In the two-layered HARM, the upper layer captures the network reachability information and the lower layer represents the vulnerability information of each node in the network. The layers of the HARM can be constructed independently of each other. This decreases the computational complexity of calculating and evaluating the HARM compared with the calculation and evaluation of the existing single-layered graphical security models.

In our prior works [12], [14], we investigated the effectiveness of defense mechanisms based on the GSM, particularly using HARM. In [12], we developed a framework to automate the security analysis of the IoT system in which HARM along with various security metrics (e.g., attack cost, attack impact) is used to assess the effectiveness of both device-level and network-level defense mechanisms in three scenarios. Similarly, in [14], we evaluated the proposed MTD techniques in a virtualized system based on HARM with a risk metric; but we analyzed the performance of three different MTD techniques, including shuffling, diversity and redundancy, separately while leaving the investigation of an integrated defense system as the future work, which is studied in this work.

Some existing approaches have adopted a risk-based or vulnerability-based security model to assess the effectiveness of defense mechanisms [1], [24], [26]. Abie and Balasingham [1] proposed a risk-based security framework for IoT environments in the eHealth domain to measure expected risk and/or potential benefits by taking a game theoretic approach and context-aware techniques. Savola et al. [26] proposed an adaptive security management scheme considering security metrics

(e.g., metrics representing authentication effectiveness, authorization metrics) to deal with the challenges in eHealth IoT environments. However, both works [1], [26] only proposed high-level ideas about the metrics without any formulation and did not consider the key characteristics of IoT environments where lightweight defense mechanisms are vital to securing a large-scale, resource-constrained IoT system. Rullo et al. [24] proposed an approach to compute the optimal security resource allocation plan for an IoT network consisting of mobile nodes and introduced a risk metric inspired by economics to evaluate the allocation plans. However, it only considered the device-level mechanisms and did not show the system-level evaluation.

Unlike the existing approaches above [1], [24], [26], we proposed a lightweight, affordable method to evaluate the deployment of an integrated defense mechanism for an IoT environment by meeting both security and performance goals of a system.

### 10.2.3 SDN Technology for IoT

**Software defined network (SDN)** is a promising technology to flexibly manage complex networks. In the SDN-based architecture, the control logic is decoupled from the switches and routers and implemented in a logically centralized controller; the controller communicates with the data forwarding devices via the southbound application programming interface (API) and provides the programmability of network applications using the northbound API. OpenFlow (OF) is the most widely used southbound API which provides the specifications for the implementation of OF switches (including the OF ports, tables, channel and protocol) [31].

Some SDN solutions are applied to IoT networks to control data flows among IoT devices [7] while others focused on applying an SDN to wireless sensor networks (WSNs) for managing sensing devices [10]. Trevizan de Oliveira et al. [7] designed and implemented an SDN architecture consisting of SDN-enabled sensor nodes and one or multiple controllers in a TinyOS environment. Galluccio et al. [10] proposed a stateful SDN solution for WSNs to reduce the amount of exchanged data between the sensor nodes and the SDN controller and to enable a variety of operations that are not supported by stateless solutions. Some SDN approaches were proposed in heterogeneous wireless networks, such as wireless access networks [17] or mobile networks [3], to manage end-to-end flows. On the other hand, other approaches provided a software-defined IoT architecture for managing IoT devices in different application domains, e.g., SD-IoT architecture for smart urban sensing in [18]. Current SDN solutions show the feasibility to control traffics within IoT networks by dynamically updating forwarding rules in either switches or SD end devices [9]. This feature could be utilized by network-level defense mechanisms to improve the response time, which can better deal with potential security issues arising from IoT networks. In our prior work [13], we designed a topology reconfiguration method for the SD WSNs with non-patchable vulnerabilities to change the attack surface of the network in order to increase the attack complexity. In this work, we consider a general IoT network with the support of SDN functionality for the network topology shuffling where an IoT network consists of both decoy and real nodes.

## 10.3 System Model

We discuss our system model in terms of (1) the network model based on SDN technology for IoT environments; (2) the attack model describing attack behaviors considered in this work; and (3) the defense model addressing defense mechanisms placed in the given network.

### 10.3.1 Network Model

In this work, we concern an IoT environment consisting of servers and IoT nodes. In a given IoT environment, nodes gather data and periodically deliver them to the servers via single or multiple hop(s) for further processing to provide a queried service. IoT nodes of different types/functionalities are placed in different Virtual Local Area Networks (VLANs) in the given network. Servers are also deployed in a separate VLAN within the network. We assume that each VLAN has a certain number of IoT nodes with similar types regarding their functionalities.

In this work, we leverage the SDN technology [7], [10], [11], [17] in order to effectively and efficiently manage and control nodes in an IoT network. There exists an SDN controller placed in a remote server. The SDN controller communicates with the SDN switches. The servers and IoT nodes in the IoT network are connected to the SDN switches. The SDN controller manages the flows between IoT nodes and servers.

### 10.3.2 Node Model

In a given network, we characterize a node's attributes in terms of four aspects: (1) whether a node is compromised or not (i.e., $n_i.c = 1$ for compromised; $n_i.c = 0$ otherwise); (2) whether a node is a real node or a decoy node (i.e., $n_i.d = 1$ for a decoy; $n_i.d = 0$ otherwise); (3) whether a node is a critical node that has confidential information where the information should not be leaked out to unauthorized parties (i.e., $n_i.r = 1$ for a critical node; $n_i.r = 0$ otherwise); and (4) a list of vulnerabilities that a node contains (i.e., $n_i.v = \{v_1, \cdots, v_m\}$ where $m$ is the total number of vulnerabilities). Hence, node $i$'s attributes are represented by:

$$A_{n_i} = [n_i.c, n_i.d, n_i.r, n_i.v] \qquad (10.1)$$

### 10.3.3 Attack Model

In this work, we consider the following attacks that may lead to breaching system security goals:

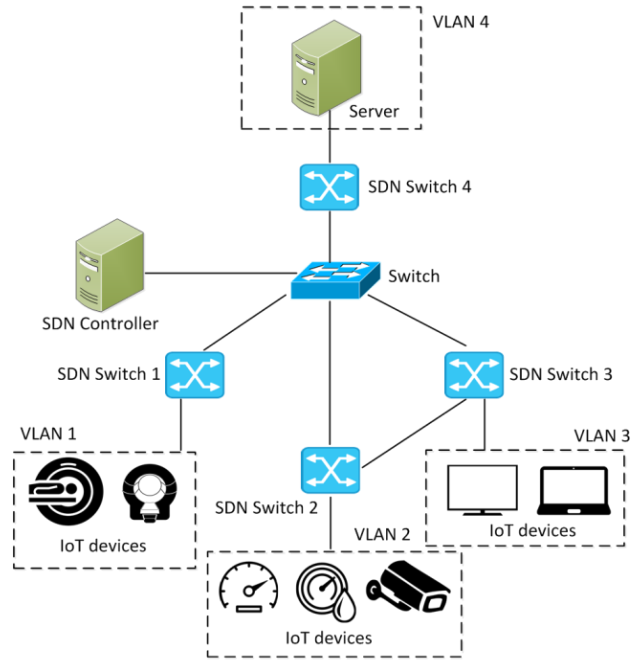

**Fig. 10.1: Example of a software-defined IoT network.**

- *Reconnaissance attacks*: Outside attackers can perform scanning attacks in order to identify vulnerable targets (e.g., a server) to break into a system (or a network). If this attack is successful, the outside attacker successfully identifies the vulnerable target and compromises it. This attack success leads to the *loss of system integrity* because the system has its system component compromised as well as the attacker from the outside. The success of this attack type is related to triggering the system failure based on the security failure condition 1 (SFC1), explained in Section 10.3.5.

- *Data exfiltration attacks*: After an outside attacker becomes an inside, legitimate attacker by using the credentials (e.g., login credentials or a legitimate key to access the system / network resources) obtained from the compromised, target node, it can leak confidential information out to unauthorized, outside parties. If this attack is successful, this results in the *loss of confidentiality* because confidential information to be only shared by legitimate users is leaked out to the unauthorized party. When this attack is successful, it will lead to the security failure based on SFC2 in Section 10.3.5.

In this work, the following attack behaviors are assumed to characterize the considered attackers:

- An attacker is assumed to have limited knowledge on whether a given node is decoy (i.e., a fake node mimicking a real, normal node) or not. The attacker's capability to detect the deception depends on the knowledge gap between the attacker and the real system state or how effectively the developed decoy system mimics the real system in a sophisticated manner. For simplicity, we use the probability of an attacker interacting with a decoy to represent the level of the attacker's intelligence in detecting a decoy node, as described in Section 10.3.4.

- After the attacker interacts with a decoy, the attacker's behavior is monitored. If the attacker realizes that the node with which it interacted is a decoy, it terminates the interactions with the decoy node immediately and attempts to find a new target to get into the system.

- An attacker's ultimate goal is to compromise the servers to leak the confidential information to unauthorized entities outside the IoT network.

- An attacker is capable of identifying exploitable, unpatched vulnerabilities or unknown vulnerabilities and compromising the vulnerable IoT nodes in a given IoT network.

- It is hard for the attacker to compromise the servers directly as each server has strong protection mechanisms. Thus, the attacker is able to exploit vulnerable IoT nodes as entry points. Once the attacker breaks into the network by using IoT nodes, it can move laterally within the network and eventually compromise the servers by identifying and exploiting unpatched or unknown vulnerabilities.

- The SDN controller is well-protected where the communications between the SDN controller and the SDN switches are assumed to be secure [11].

### 10.3.4 Defense Model

We assume that traditional defense mechanisms are in place on the IoT network, including a network-based IDS, firewalls, and anti-virus software on the servers. The IDS is capable of monitoring the whole IoT network and creates alerts once an intrusion is being detected for incident response. In addition, we have two types of intrusion prevention mechanisms in place, cyberdeception and MTD, to dynamically change the attack surface of the IoT network so that they can make attackers hard to launch their attacks, resulting in high attack complexity.

**10.3.4.1 Decoy System as Defensive Cyberdeception**

This defense mechanism allows defenders to capture and analyze malicious behaviors by luring attackers into a decoy system within a given network and interacting with the attackers. The decoy system is deployed independently from the real system. Accordingly, we assume that normal, legitimate users are not aware of the existence of the decoy system while the defenders will only get alerts caused by the malicious intrusions if an attacker breaks into the decoy system.

We consider two types of decoys utilized throughout an IoT network considered in this work:

- *Emulation-based decoys*: This type of decoys allows defenders to create a variety of fake assets and to provide a large-scale coverage across the network.

- *Full OS-based decoys*: This type of decoys enables the replication of actual production devices to increase the possibility that the attacker engages the decoys and exposes its intention and/or strategy.

Both emulation-based and full OS-based decoys can be autonomously created to fit within the environment with no changes to the existing infrastructure. They can provide various types of interactive capabilities. To increase the overall chances of accessing decoys by attackers, the decoys can be created by combining multiple, diverse forms of decoys that look like real, legitimate nodes. In addition, the decoys can be deployed in every VLAN of the network. Besides, there exists an intelligence center performing the following tasks: (1) create, deploy, and update a distributed decoy system; (2) provide automated attack analysis, vulnerability assessment, and forensic reporting; and (3) integrate the decoy system with other prevention systems (e.g., security incident and event management platform, firewalls) to block attacks. The module for the decoy node deployment can be implemented and placed in a remote server.

For these two types of decoy nodes to be considered in this work, we create a design parameter, $P_d$, indicating the probability that an attacker interacts with a decoy node. Since full-OS-based decoys are considered as having more sophisticated, real-node like quality with more cost, we consider $P_d^{em}$ as the probability that the attacker interacts with an emulation-based decoy while using $P_d^{OS}$ as the probability based on the assumption that the attacker will more likely to interact with an OS-based decoy with $P_d^{em} \leq P_d^{OS}$.

**10.3.4.2 Network Topology Shuffling-based MTD**

In this work, we consider Network Topology Shuffling-based MTD, namely NTS-MTD, to change the topology of the given IoT network where the network consists of both real, legitimate nodes and decoy nodes. NTS-MTD is to be triggered following the concept of event-based MTD in that the network topology changes upon a certain event indicating that the network is at risk due to too many nodes being compromised. For example, following the concept of Byzantine Failure [11], if the system is close to a security failure state, such as more than one third of nodes being compromised, the network topology is being shuffled in order to stop or mitigate the spread of nodes being compromised.

In this work, we assume that the SDN controller can control the traffic of the decoy nodes in an SDN-based decoy system and manage to change the network topology upon a certain event. We combine the cyberdeception and NTS-MTD and propose a network topology shuffling with decoy nodes to change the attack surface of the IoT network. The details of the proposed decoy system and the NTS-MTD are described in Section 10.4.

### 10.3.5 Security Failure Conditions

A system fails when either of the below two conditions is met:

- **Security Failure Condition 1 (SFC1)**: This system failure is closely related to the attacker's successful reconnaissance attacks and accordingly their successful compromise of system components (or capture of IoT nodes). We define this system failure based on the concept of Byzantine Failure [11]. That is, when more than one third of legitimate, member nodes are compromised, we define it as the system failure due to the *loss of system integrity*.

- **Security Failure Condition 2 (SFC2)**: This system failure occurs when confidential information is leaked out to unauthorized parties by inside attackers (or compromised nodes), which perform the so called *data exfiltration attack*. This type of system failure occurs due to the *loss of data confidentiality*.

## 10.4 Proposed Proactive Defense Mechanisms

In this section, we provide the overview of our proposed NTS-MTD in terms of the decoy-based network topology shuffling, optimization method, and graphical security model for security analysis.

### 10.4.1 Network Topology Shuffling with Decoy Nodes

In this section, we describe the details of the proposed NTS-MTD in terms of how to deploy the initial set of nodes, how to select a network topology to be shuffled, and when to shuffle a network topology.

#### 10.4.1.1 Initial Deployment of Nodes

We consider both servers and IoT node decoys to be deployed in the IoT network. As the network is divided into different VLANs, a certain number of decoy nodes can be placed into each VLAN based on the real nodes placed to the corresponding

VLAN. At least one decoy server needs to be deployed to interact with the attacker and reveal the attacker's intent. We distribute the IoT decoy nodes into each VLAN based on the deployment of real nodes in the VLAN. Note that we can deploy more decoys if the VLAN has a large number of real nodes with different types. When adding the decoy nodes, we link real IoT nodes with the decoy nodes to lure attackers into the decoy system. The flows from the real IoT nodes to the decoy nodes or from the decoy nodes to the decoy nodes are controlled by the SDN controller. There will be no flows from decoy nodes to real nodes as the decoy nodes are used to divert the attackers from the real system; once the attacker is lured into the decoy system, it will be diverted to other decoys within the decoy system and the behavior will be monitored; if the attacker finds out a node it interacted with is a decoy node, it will terminate the interaction with the decoy node and look for a new target to break in. The directional flows between real and decoy nodes may reveal some information to the attackers in the long term. We will consider changing flows from real nodes to real nodes in the future work to increase the complexity of the connection changes.

We assume that newly added flows will not affect normal flows from the IoT nodes to servers for the service delivery. In practice, the IoT nodes will consume more energy to deliver more flows and may delay the time to send normal packets towards the server. Thus, the service availability provided by the server may be affected. This will be examined in our future work.

### 10.4.1.2 Selection of a Network Topology to Shuffle

Given the IoT network with real nodes, we first decide the number of decoys to be potentially deployed in each VLAN and then randomly generate a set of deployments of the decoy nodes with added connections to some randomly chosen real nodes. The set of these randomly generated network topologies will be used as the initial set of topologies and they will be used in the shuffling optimization algorithm to identify an optimal network topology.

### 10.4.1.3 Adaptive Shuffling of a Network Topology as an MTD

As a baseline MTD, we can simply choose a fixed time interval to be used to execute a given MTD mechanism. Alternatively, we can design an intelligent MTD to be executed in an *adaptive* manner based on the system security level detected by the defender. For example, a network topology needs to be changed when the defender perceives the system security vulnerability level larger than a given threshold $\rho$. The system security vulnerability level at time $t$, denoted by $SSV(t)$, is measured by two dimensions: (1) how many legitimate, inside nodes are compromised until time $t$, associated with SFC1; and (2) how many neighboring nodes of a critical node (i.e., $n_i.r = 1$) within $k$ hops from the critical node $i$ are compromised until time $t$, which is related to detecting SFC2. Note that when the system meets either SFC1 or SFC2, then the system fails, leading to $SSV(t) = 1$. Otherwise, $SSV(t)$ is computed by:

$$SSV(t) = w_1 \frac{CN(t)}{N} + w_2 \frac{CN_{ck}(t)}{N_{ck}(t)} \tag{10.2}$$

Here $w_1$ and $w_2$ are weights to consider SFC1 and SFC2, respectively, where $w_1 + w_2 = 1$. $N$ is the total number of real nodes initially deployed and $CN(t)$ is the number of compromised, real nodes at time $t$. $N_{ck}(t)$ is the total number of real nodes within $k$ hops from given critical nodes at time $t$ while $CN_{ck}(t)$ is the total number of compromised, real nodes among the real nodes within the $k$ hops from the given critical nodes. Since there may be multiple critical nodes which have confidential information that should not be leaked out to outside, non-authorized nodes, we estimate $CN_{ck}(t)$ by:

$$CN_{ck}(t) = \sum_{i \in L_k(t)} n_i.c\,(t) \tag{10.3}$$

where $L_k(t)$ means the number of real nodes that belong to neighbors of any critical nodes within $k$ hops from them at time $t$. Recall that $n_i.c(t)$ refers to whether node $i$ is compromised ($n_i.c(t) = 1$) or not ($n_i.c(t) = 0$) at time $t$. Thus, the cardinality of $L_k(t)$ (i.e., $|L_k(t)|$) is the same as $N_{ck}(t)$. Note that as a network topology keeps changing due to executing the MTD to change the network topology, both $N_{ck}(t)$ and $CN_{ck}(t)$ are the functions of time to reflect their dynamic changes. Note that the set $L_k(t)$ may include any critical nodes being compromised. If this happens, it means the system meets SFC2 and the system failed. That is, $SSV(t) = 1$ and no further detection of system security level is needed.

For each $SSV(t)$, we calculate the mean time to compromise (MTTC) associated with it. MTTC refers to the total amount of time that the attacker takes to compromise a series of nodes within the network until the system reaches a certain security vulnerability level. The computation of MTTC is detailed in Section 10.5.3.

### 10.4.2 Genetic Algorithm-based Network Shuffling Optimization

In this section, we discuss a GA-based network shuffling optimization technique in a given SDN-based IoT network.

To breach system security goals, an attacker may be able to find multiple attack paths to reach a target node via one or multiple entry points. An attack path specifies a sequence of nodes that the attacker can compromise to reach the target node. We consider a set of attack paths $AP$ for the attacker to reach all the targets from all possible entry points. Each attack path $ap$ is a sequence of nodes over the attack path. We use $AP_r$ to represent a set of attack paths with real nodes as targets and $AP_d$ to indicate a set of attack paths with decoy nodes as targets. $AP_r$ only contains the real nodes while $AP_d$ contains both real nodes and decoy nodes. To be specific, in the attack model, we assume the attacker could exploit any IoT nodes as entry points. Once the attacker successfully compromises a node, it could use this node as the stepping stone to compromise other nodes and further compromise servers as their targets. The attacker may find a real node as the entry point and then is diverted to a decoy node. Once the attacker is lured into the decoy system, it will be diverted to other decoy nodes within the decoy system. If the attacker reaches the decoy server, this is accounted as an attack path in $AP_d$; but if the attacker figures out the decoy node and terminates its interaction with the decoy node, this is not counted as an attack path because the attacker does not reach the decoy server (i.e., target). Besides, the decoy nodes could be easily updated or cleared once being compromised via the central management portal thus the attacker will not recognize the same decoy node during the subsequent attacks.

We design three metrics to be optimized: (1) the number of attack paths towards the decoy targets ($N_{DT}^{AP}$); (2) mean time to security failure (MTTSF); and (3) defense cost ($C_D$). The computations of these metrics are described in Section 10.5.2.

### 10.4.3 Graphical Security Model for the IoT MTD

We apply the graphical security model to assess the security of an SDN-based IoT network with the proposed proactive defense mechanism. Fig. 10.2 describes the workflow consisting of network generation, topology generation, security model generation, shuffling mechanism evaluation, and shuffling optimization.

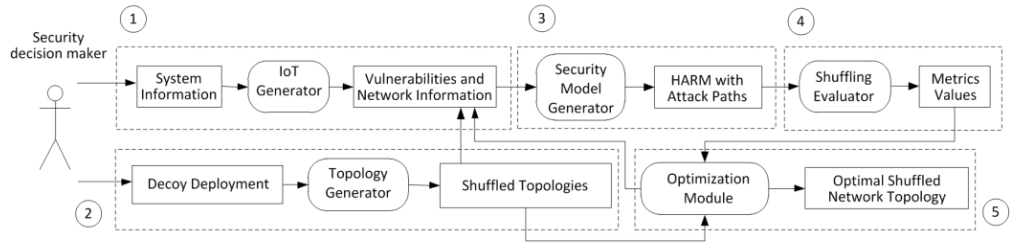

**Fig. 10.2: Workflow for security analysis.**

The workflow of this security analysis consists of the following five phases:

- **Phase 1**: The security decision maker provides the `IoT Generator` with the system information (i.e., an initial network topology and node vulnerability) to construct an IoT network.

- **Phase 2**: Given the network and initial deployment of the decoys, the `Topology Generator` randomly generates a set of different topologies (i.e., add connections from real nodes to decoys) based on the shuffling algorithm. Each shuffling is presented in an integer format and passed onto the `Optimization Module`. **Phase 3**: The `Security Model Generator` takes the shuffled network as input and automatically generates the HARM which captures all possible attack paths. We use the three-layered HARM [14], [15] as our graphical security model. In the three-layered HARM, the upper layer captures the subnet reachability information, the middle layer represents the node connectivity information (i.e., nodes connected in the topological structure), and the lower layer denotes the vulnerability information of each node.
- **Phase 4**: The `Shuffling Evaluator` takes the HARM as input along with the evaluation metrics and computes the results which are then fed into the `Optimization Module`.
- **Phase 5**: Based on the initial set of shuffled topologies and the associated evaluation results, the `Optimization Module` applies the multi-objective genetic algorithm to compute the optimal topology for the IoT network based on the termination conditions (e.g., the maximum number of generations defined by the security decision maker).

## 10.5 Numerical Results & Analysis

In this section, we describe our experimental setup and simulation results, along with the analysis of the observed trends from the obtained results.

### 10.5.1 Experimental Setup

We use the example SD-IoT network shown in Fig. 10.1. Specifically, the network consists of four VLANs. There are two Internet of Medical Things (i.e., MRI and CT Scan) in $VLAN_1$, a smart thermostat, a smart meter and a smart camera in $VLAN_2$, a smart TV and a laptop in $VLAN_3$ and a server located in $VLAN_4$. $VLAN_4$ is connected with other three VLANs as IoT devices need to deliver information to the server for further processing. $VLAN_2$ is also connected to $VLAN_3$ for the applications on smart TV and laptop to control the smart sensors as well as receiving videos from the smart camera.

For the software vulnerability analysis, we will be using Common Vulnerabilities and Exposures (CVE) / National Vulnerability Database (NVD) [21] for IoT networks. We assume each real node has one vulnerability that could be exploited by the attacker to gain a root privilege. More vulnerabilities could be chosen for the nodes in the future work as the focus of the research is to propose and evaluate the proactive defense mechanism, rather than demonstrating the capabilities of the graphical security model to analyze the security posture of the IoT network with multiple vulnerabilities. The vulnerability information of real nodes (i.e., CVE ID) are assumed following Table 10.1.

**Table 10.1: Real node and vulnerability information.**

| Real Node | VLAN | CVE ID | Compromise Rate |
|---|---|---|---|
| MRI | VLAN1 | CVE-2018-8308 | 0.006 |
| CT Scan | VLAN1 | CVE-2018-8308 | 0.006 |
| Smart Thermostat | VLAN2 | CVE-2018-11315 | 0.006 |
| Smart Meter | VLAN2 | CVE-2017-9944 | 0.042 |
| Smart Camera | VLAN2 | CVE-2018-10660 | 0.042 |
| Smart TV | VLAN3 | CVE-2018-4094 | 0.012 |
| Laptop | VLAN3 | CVE-2018-8345 | 0.004 |
| Server | VLAN4 | CVE-2018-8273 | 0.006 |

**Table 10.2: Decoy node and vulnerability information.**

| Decoy Node | VLAN | CVE ID | Compromise Rate |
|---|---|---|---|
| CT Scan | VLAN1 | CVE-2018-8308 | 0.006 |
| | | CVE-2018-8136 | 0.012 |
| Smart Camera | VLAN2 | CVE-2018-6294 | 0.042 |
| | | CVE-2018-6295 | 0.042 |
| | | CVE-2018-6297 | 0.042 |
| Smart TV | VLAN3 | CVE-2018-4094 | 0.012 |
| | | CVE-2018-4095 | 0.012 |
| Server | VLAN4 | CVE-2016-1930 | 0.042 |
| | | CVE-2016-1935 | 0.012 |
| | | CVE-2016-1962 | 0.042 |

We also assume the compromise rate of each vulnerability. The compromise rate represents the frequency that an attacker could successfully exploit the vulnerability to gain root privilege per time unit (i.e., hour). We estimate the value according to the base score from the Common Vulnerability Scoring System (CVSS). Specifically, we estimate the compromise rate as once per day (i.e., 0.042) if the base score is 10.0, twice per week (i.e., 0.012) with the base score of around 8.0, once per week (0.006) when the base score is around 7.0 and once per 10 days (i.e., 0.004) under the score of around 5.0. This value will be used to calculate MTTSF and MTTC in the simulations based on the HARM.

We consider using one type of decoys in each VLAN in the initial deployment of the decoy system. In order to lure the attackers, each decoy is assumed to be configured to have multiple vulnerabilities. The attacker could exploit any vulnerability to gain the root permission of the node. We assume to use the vulnerabilities of the decoys based on Table 10.2.

## 10.5.2 Metrics

We use the following metrics to measure security and performance of the proposed proactive defense mechanism:

- **Number of attack paths towards decoy targets ($N_{DN}^{AP}$)**: This metric indicates the level of deception that diverts the attacker from the real system. $N_{DT}^{AP}$ is calculated by $|AP_d|$ to sum the attack paths towards the decoy targets.
- **Mean Time To Security Failure (MTTSF)**: This metric measures the system lifetime indicating how long the system prolongs until the system reaches either SFC1 or SFC2 (described in Section 10.3.5). That is, this measures the system uptime without occurring any security failure. MTTSF is measured by:

$$MTTSF = \sum_{i \in S}(1 - SF_i)\int_{t=0}^{\infty} P_i(t)dt \qquad (10.4)$$

where $S$ is a set of all system states and $SF_i$ returns 1 when system state $i$ reaches either SFC1 or SFC2; 0 otherwise. $P_i(t)$ indicates the probability that the system is at system state $i$.

- **Mean Time To Compromise (MTTC):** This metric refers to the total amount of time that the attacker takes to compromise a series of nodes within the network until the system reaches a certain security vulnerability level, $SSV$. This metric is used to detect the system security vulnerability level upon the number of nodes being compromised by the attacker and employed to determine when to trigger an MTD operation. Similar to the computation of MTTSF as above, MTTC is estimated by:

$$MTTC = \sum_{i \in S} S_i \int_{t=0}^{\infty} P_i(t)dt \qquad (10.5)$$

where $S$ refers to a set of all system states that do not reach the given $SSV$ and $S_i$ returns 1 when system state $i$ didn't reach the $SSV$; 0 otherwise. $P_i(t)$ is the probability of the system being at system state $i$.

- **Defense Cost ($C_D$):** This metric captures the cost associated with the shuffling operations. That is, we count the number of edges shuffled (i.e., from connected to disconnected or from disconnected to connected) by:

$$C_D = \int_{t=0}^{MTTSF} C_S(t) \qquad (10.6)$$

where $C_S(t)$ refers to the number of shuffled edges at time $t$. Note that a same edge can be shuffled multiple times over time and each shuffling is counted as a separate MTD operation during the system uptime.

### 10.5.3 Comparing Schemes

We have two aspects of MTD to investigate: (i) when to shuffle a network topology; and (ii) how to select the network topology. As described in Section 10.4.2, (i) is to investigate an interval or when to execute the proposed MTD operation in an adaptive manner while (ii) is to investigate the effect of a selected network topology generated by a GA-based network topology or a random network topology.

The two types of *'when to shuffle a network topology' strategies* are:

- **Fixed Shuffling (FS)**: This shuffling represents a baseline scheme using a fixed time interval to shuffle a given network topology.

- **Adaptive Shuffling (AS)**: This shuffling is designed to execute the MTD in an *adaptive* manner based on the system security vulnerability level ($SSV(t)$), detected by the defender with two given thresholds: (1) $\beta$ to check the decrease of the $SSV$ during the time $\Delta$; and (2) $\rho$ to check the current system security vulnerability, $SSV(t)$, as described in Section 10.4.1. The NTS-MTD is executed if the condition, $(SSV(t) - SSV(t - \Delta) > \beta) \wedge (SSV(t) < \rho)$, is met where $\Delta$ is a checking interval.

The two types of *'how to select a network topology' strategies* are:

- **Random Network Topology (RNT)**: This scheme is used as a baseline model that can simply select a network topology based on a simple random selection of a node's edges to other nodes based on a rewiring probability $P_r$ which represents the probability that a node is connected with another node in a given network. Here $P_r$ is critical in determining the overall network density in a given network.

- **GA-based Network Topology (GANT)**: This scheme selects a network topology based on the method proposed in Section 10.4.1. That is, a network topology to be used for a next round of shuffling is selected based on the network topology that maximizes the objective functions used in the GA, as discussed in Section 10.4.1.

Since we have two strategies under each category of the proposed MTD mechanisms, we investigate the following four schemes as follows:

- **FS-RNT:** This scheme executes the NTS-MTD based on a certain fixed time interval, $\gamma$, with a selection of a random network topology based on RNT.

- **AS-RNT:** This scheme adaptively executes the NTS-MTD based on the level of system security vulnerability, $SSV(t)$, detected by the defender selecting a random network topology based on RNT.
- **FS-GANT:** This scheme changes a current network topology to a selected network topology based on the decision of the fitness functions using GA and executes the NTS-MTD based on a certain fixed time interval.
- **AS-GANT:** This scheme adaptively changes a network topology based on the system security vulnerability level, $SSV(t)$, detected by the defender and selects a network topology to change based on the decision of the fitness functions using GA.

### 10.5.4 Simulation Steps and Parameter Details

We implement the simulations based on the workflow shown in Figure 10.2. The `Shuffling Evaluator` could be either a GA-based shuffling algorithm or a random shuffling algorithm.

We consider MTTSF as an expected mean MTTSF, $E(\overline{MTTSF})$, in computing the shuffled network topology based on the assumptions of the potential attacker (See Section 10.3.3). For each shuffled network, we construct HARM and calculate potential attack paths. We assume that the attacker randomly selects one entry point and compromises the nodes on the attack path at each time until either security failure condition (see Section 10.3.5) is met. We assume the defender will clear the decoy nodes once it detects the attacker's interaction with the decoy target. Therefore, the attacker will not recognize the same decoy node during its following action. The attacker's intelligence, estimated by the probability to interact with the decoy, $P_d^{em}$ and $P_d^{OS}$ (see Section 10.3.4) is incorporated into the calculation of the expected mean MTTSF, $E(\overline{MTTSF})$, as well as actual MTTC in adaptive shuffling. We consider the defender's error probabilities in estimating the attacker's interaction probability with a decoy with the ranges of $[-0.05, 0.05]$. This affects the defender's ability to estimate $E(\overline{MTTSF})$ because the defender needs to compute the actual MTTC based on its detection probability, $\alpha = 0.95$ (i.e., a defender's confidence about the attacker's intelligence). The detailed calculation steps for these two metrics are found in [12]. As the entry points are randomly chosen, we run the attacker model for 100 times and calculate $E(\overline{MTTSF})$ from the HARM for the given shuffled network.

In GANT, we encode each shuffling solution for the whole network as a binary valued vector where 1 represents the existence of an edge between two nodes (i.e., two nodes are connected) while 0 represents no edges (i.e., two nodes are not connected). We limit the potential connections to be the edges from real IoT nodes to decoy nodes. Hence, to optimize the defense cost, we aim to maximize $C_T(t) -$

$C_D(t)$ where $C_T(t)$ refers to the total defense cost that counts the total number of potential changes of the edges at time $t$ while $C_D(t)$ is the number of edges changed by the used NTS-MTD (see Section 10.5.2) at time $t$. Here we aim to solve a multi-objective optimization (MOO) problem with three objectives to maximize $N_{DN}^{AP}$ and $E(\overline{MTTSF})$ while minimizing $C_D$ (or maximizing $C_T(t) - C_D(t)$). The optimization problem is to compute a set of Pareto optimal solutions (or Pareto frontier) [6]. In order to choose one optimal solution among the Pareto frontier, we first normalize the three metrics, denoted by $\widetilde{N_{DN}^{AP}}$, $E(\widetilde{\overline{MTTSF}})$, and $\widetilde{C_D}$, and assign a weight for each metric based on scalarization-based MOO technique to make the MOO problem to a single-objective optimization (SOO) problem [6], respectively. The metric is normalized by:

$$\tilde{X} = \frac{X}{X_{max}} \qquad (10.7)$$

where $\tilde{X}$ is the normalized metric value, $X$ is the original metric value, and $X_{max}$ is the maximum metric value of the corresponding fitness function among the final population in the GA-based algorithm.

The objective function we aim to maximize is represented by:

$$\max \; w_N \widetilde{N_{DN}^{AP}} + w_M E(\widetilde{\overline{MTTSF}}) + w_C \widetilde{C_D} \qquad (10.8)$$

where $w_N$, $w_M$ and $w_C$ are weights for each metric with $w_N + w_M + w_C = 1$. The optimal solution is the network topology with the maximum objective value.

In this study, we consider an equal weight for $w_N, w_M$ and $w_C$, respectively, with 1.0/3.0.

We assume the following algorithm parameters for the simulations with GANT: population size ($N$) = 100, maximum number of generations ($N_g$) = 100, crossover rate ($r_c$) = 0.8 and mutation rate ($r_m$) = 0.2. In RNT, we randomly change the edges between the real IoT nodes and decoy nodes. We also considered the probability that an edge will be shuffled (i.e., add/remove an edge) with $P_r = 0.5$.

When calculating $SSV(t)$ for adaptive shuffling, we use the following default values for the weights ($w_1, w_2$), two $SSV$ related thresholds to execute the NTS-MTD ($\beta$ and $\rho$) and the $k$ number of hops to determine the neighbor nodes to a critical node: $w_1 = w_2 = 0.5$, $\beta = 0.01$, $\rho = 0.1$ and $k = 1$. The checking interval $\Delta$ is determined upon detecting a compromised node by the defender.

**Table 10.3: Design parameters, their meanings, and the default values.**

| Parameter | Meaning | Default Value |
|---|---|---|
| $N$ | The total number of network topologies with initial decoy deployment and randomly generated connections between real and decoy nodes | 100 |
| $N_{DN}^{AP}$ | The number of attack paths towards the decoy targets | Metric |
| MTTSF | Mean time to security failure, representing the system lifetime | Metric |
| MTTC | Mean time to compromise a fraction of nodes in a network until the system reaches a certain security vulnerability level | Metric |
| $C_D$ | The number of edges changed from a previous network topology to a current network topology due to the network shuffling-based MTD | Metric |
| $w_N$ | A weight to consider $N_{DN}^{AP}$ | 1/3 |
| $w_M$ | A weight to consider MTTSF | 1/3 |
| $w_C$ | A weight to consider $C_D$ | 1/3 |
| $w_1$ | A weight to consider the security vulnerability associated with SFC1 | 0.5 |
| $w_2$ | A weight to consider the security vulnerability associated with SFC2 | 0.5 |
| $N_g$ | The maximum number of the generation used in GANT | 100 |
| $r_c$ | Crossover rate used in GANT | 0.8 |
| $r_m$ | Mutation rate used in GANT | 0.2 |
| $P_r$ | The probability of an edge being shuffled | 0.5 |
| $\beta$ | The threshold used to estimate the decrease of the system security vulnerability level during the time $\Delta$ used in GANT | 0.01 |
| $\rho$ | The threshold of tolerating system security vulnerability used in GANT | 0.1 |
| $k$ | The number of hops to determine a node's ego network | 1 |
| $\gamma$ | The fixed shuffling time interval used in RNT (hour) | 24 |

Whenever the defender detects a compromised real node, it will check whether the system reaches the given $SSV(t)$, reflecting the nature of an event-driven adaptive MTD.

We assume there is an attacker during each simulation run. The attacker randomly chooses entry points and compromises nodes along the attack paths with the behaviors defined in Section 10.3.3. By using the fixed shuffling schemes, the network may be shuffled while a node is under attacks. We assume the attacker is forced to quit the network due to lost connections and needs to find other ways to break into the network. During the subsequent attack after shuffling, the attacker could continue its previous attack action once it encounters the same real node next time (i.e., MTTC for the real node is accumulated throughout the MTTSF). By using the adaptive shuffling schemes, the network is shuffled due to the system security vulnerability level detected by the defender. The attacker is also forced to quit the network after each shuffling due to lost connections and needs to find ways to re-enter the network. For both schemes, the decoy node is cleared at each shuffling.

For each simulation, we calculate actual MTTSF under the existence of real attackers, average $N_{DT}^{AP}$, denoted as $\overline{N_{DT}^{AP}}$, by summing $N_{DT}^{AP}$ from each shuffled network and dividing it by the total number of shuffling times within MTTSF, and defense cost per time unit, $\widehat{C_D}$, by summing $C_D$ from each shuffling network and dividing it by actual MTTSF.

We run the entire simulation for 100 times and calculate the average metric value by summing the metric value from each simulation and dividing the summed value by the total number of simulations. We use $\overline{\overline{MTTSF}}$, $\overline{\overline{N_{DT}^{AP}}}$ and $\overline{\widehat{C_D}}$ to represent the mean metric value in our experimental results shown in the following section. Table 10.3 summarizes the design parameters, their meanings, and corresponding default values used in our simulation experiments.

### 10.5.5 Simulation Results

In our simulation experiments, we consider three scenarios by varying the number of decoys in each VLAN and the level of attackers' intelligence in detecting decoy nodes (i.e., $P_d^{em}$ and $P_d^{OS}$): (1) one decoy node assigned for each VLAN in the presence of low-intelligent attackers (i.e., $P_d^{em} = 0.9$ and $P_d^{OS} = 1.0$); (2) one decoy node assigned for each VLAN in the presence of medium-intelligent attackers (i.e., $P_d^{em} = 0.5$ and $P_d^{OS} = 0.9$); and (3) two decoy nodes assigned for each VLAN in the presence of low-intelligent attackers (i.e., $P_d^{em} = 0.9$ and $P_d^{OS} = 1.0$). For other key design parameters, we will follow default values summarized in Table 10.3.

#### 10.5.5.1 One Decoy Node for Each VLAN with Low-Intelligent Attackers

Fig. 10.3 shows how the considered four MTD schemes perform in terms of mean values of the three metrics described in Section 10.5.2 over 100 simulations. In Fig. 10.3 (a), with the random shuffling, both FS-RNT and GA-RNT have a similar number of attack paths towards decoy targets, 29; with the GA-based shuffling, $\overline{\overline{N_{DT}^{AP}}}$ is much higher with 54 for both schemes. $\overline{\overline{N_{DT}^{AP}}}$ is determined by the network topology; thus, either random shuffling or GA-based schemes have a similar performance in $\overline{\overline{N_{DT}^{AP}}}$.

Fig. 10.3 (b) shows the values of the average MTTSF. With FS, both schemes have higher $\overline{MTTSF}$ compared with AS schemes based on the following reason. A node may still be under attacks while the network is shuffled because the fixed interval is much smaller than MTTC for the node. However, the attacker is forced to quit the network due to lost connections and needs to re-enter the network by randomly choosing entry points to compromise.

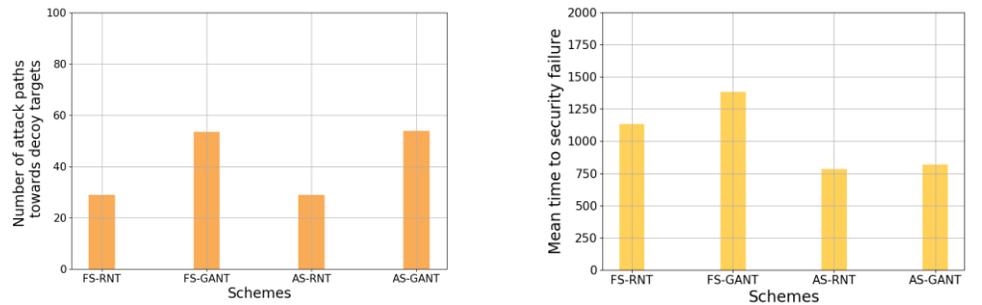

(a) Average number of attack paths towards decoy targets ($\overline{\overline{N_{DT}^{AP}}}$)

(b) Average MTTSF ($\overline{MTTSF}$)

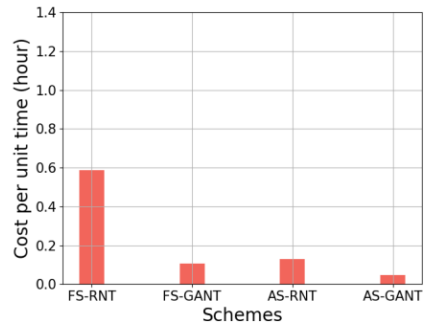

(c) Average defense cost per time unit ($\overline{\overline{C_D}}$)

**Fig. 10.3: Performance comparison of the four MTD schemes with low-intelligent attackers where one decoy node is deployed for each VLAN.**

During the subsequent attack after each shuffling, the attacker could continue its previous attack action once it encounters the same real node next time (i.e., MTTC for the real node is accumulated throughout MTTSF). After then, the attacker needs to launch a new attack for decoys as they are cleared at each shuffling. However, although MTTC is accumulated for the real node, the time to meet either security failure condition (i.e., compromise a certain number of nodes or the critical node) increases over time. In addition, FS-GANT has higher $\overline{MTTSF}$ (i.e., 1,381 hours) than that of FS-RNT (i.e., 1131 hours) while AS-GANT also has slightly higher $\overline{MTTSF}$ than that of AS-RNT, showing 782 and 817 hours, respectively.

Now we investigate the effect of the four schemes in terms of the average defense cost per time unit which shows the tradeoff for defense cost per time unit and MTTSF. Fig. 10.3 (c) shows the average defense cost per time unit $\overline{\overline{C_D}}$ under the four schemes. As expected, FS-RNT scheme incurs the highest cost (i.e., 0.59) due to frequent executions of network shuffling while AS-GANT has the lowest cost (i.e., 0.05). Surprisingly, FS-GANT incurs less cost than AS-RNT with 0.11 and 0.13, respectively. This is due to less edge changes of GA-based scheme in each shuffling even if the network topology changes at a fixed interval.

Overall, GA-based schemes can preserve a higher security level in terms of $N_{DN}^{AP}$ and maintaining a lower cost while fixed shuffling-based schemes incur higher $\overline{MTTSF}$. We can see there is a balance between MTTSF and defense cost per time unit. $\overline{MTTSF}$ of FS-GANT is roughly 1.7 times higher than $\overline{MTTSF}$ of AS-GANT while $\overline{\overline{C_D}}$ of FS-GANT is twice of that of AS-GANT.

### 10.5.5.2 One Decoy Node for Each VLAN with Medium-Intelligent Attacker

We use the same initial decoy deployment in Section 10.5.5.1, except considering medium-intelligent attackers with $P_d^{em} = 0.5$ and $P_d^{OS} = 0.9$. For other design parameters, we follow their default values summarized in Table 10.3.

Fig. 10.4 shows the performance of the four schemes in terms of the mean value of each metric in Section 10.5.2 based on 100 simulation runs when the attackers have the medium-intelligent levels. Fig. 10.4 (a) shows the similar trend observed in Fig. 10.3 (a). With the random shuffling, both FS-RNT and GA-RNT have a similar number of attack paths towards decoy targets which is about 29; for the GA-based shuffling, much higher $\overline{N_{DT}^{AP}}$ is observed, showing 54 for both schemes.

Fig. 10.4 (b) shows a similar trend as Fig. 10.3 (b). With FS, both schemes have higher $\overline{MTTSF}$ compared with AS schemes. Besides, FS-GANT has higher $\overline{MTTSF}$ (i.e., 1,414 hours) than FS-RNT (i.e., 1,139 hours) while AS-GANT also has slightly higher $\overline{MTTSF}$ than AS-RNT, showing 720 and 766 hours, respectively.

Fig. 10.4 (c) shows the similar trend in Fig. 10.3 (c). As expected, FS-RNT incurs the highest cost (i.e., 0.59) among all due to frequent executions of network shuffling while AS-GANT has the lowest cost (i.e., 0.05). Surprisingly, FS-GANT incurs less cost than AS-RNT, showing 0.1 and 0.14, respectively, due to less edge changes of GA-based scheme in each shuffling.

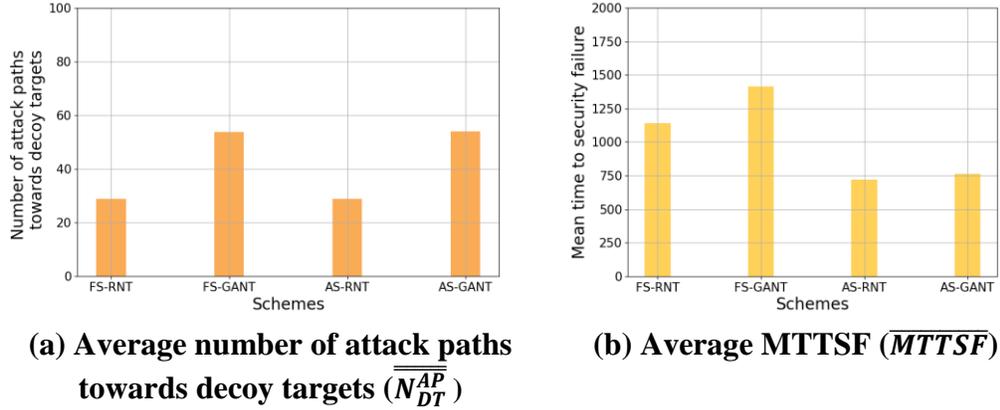

(a) Average number of attack paths towards decoy targets ($\overline{\overline{N_{DT}^{AP}}}$)

(b) Average MTTSF ($\overline{MTTSF}$)

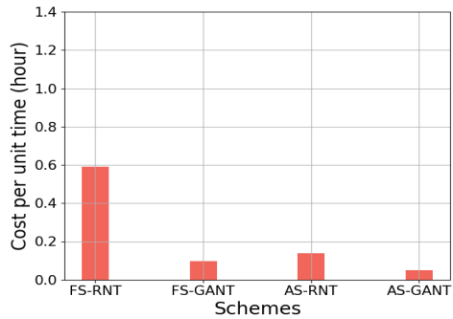

(c) Average defense cost per time unit ($\overline{\overline{C_D}}$)

**Fig. 10.4: Performance comparison of the four MTD schemes with medium-intelligent attackers where one decoy is deployed for each VLAN.**

Overall, we observe GA-based schemes can maintain a higher security level in terms of $N_{DN}^{AP}$ and generate lower cost while fixed shuffling schemes incur higher $\overline{MTTSF}$. We can also see there is a balance between MTTSF and defense cost per time unit. $\overline{MTTSF}$ of FS-GANT is roughly 1.8 times higher than that of AS-GANT while $\overline{\overline{C_D}}$ of FS-GANT is twice of that of AS-GANT. Additionally, fixed shuffling-based schemes could keep $\overline{MTTSF}$ similar under low and medium attack intelligence. This is because the fixed interval is much smaller than MTTC for a node. Therefore, a node may still be under attacks while the network is shuffled. However, the attacker is forced to quit the network and could continue its previous attack when encountering the same real node thus causing MTTC for the real node being accumulated throughout the MTTSF. Although the attacker's intelligence increases causing shorter MTTC for decoy nodes, the time to meet either security failure condition (i.e., compromise a certain number of nodes or the critical node) remains similar. For adaptive shuffling-based schemes, $\overline{MTTSF}$ with low-intelligent attackers is slightly higher than that with medium-intelligent attackers. This implies that potential attackers with higher intelligence in detecting the decoys hurts the system lifetime as measured based on MTTSF under adaptive shuffling. However, we prove that AS-GANT is resilient under high-intelligent attacks without much reduction of MTTSF when compared with the case with low-intelligent attacks as shown in Section 10.5.5.1.

### 10 5.5.3 Two Decoy Nodes for Each VLAN with Non-Intelligent Attackers

We use two decoy nodes for each VLAN as the initial deployment. We use the low-intelligent attackers and follow the default values of other design parameters in Table 10.3.

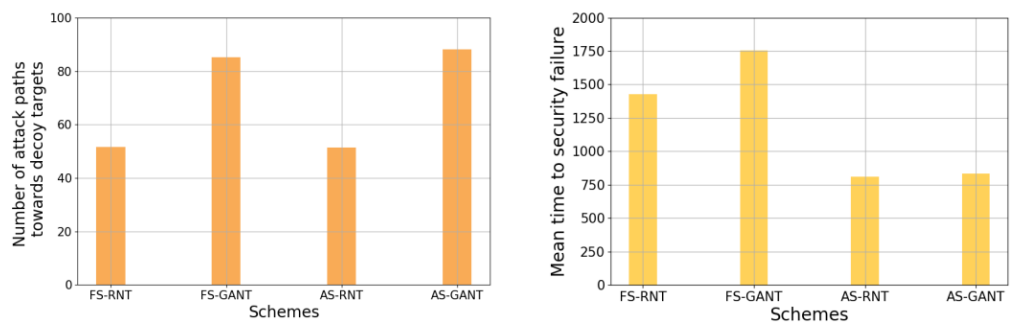

(a) Average number of attack paths towards decoy targets ($\overline{N_{DT}^{AP}}$)

(b) Average MTTSF ($\overline{MTTSF}$)

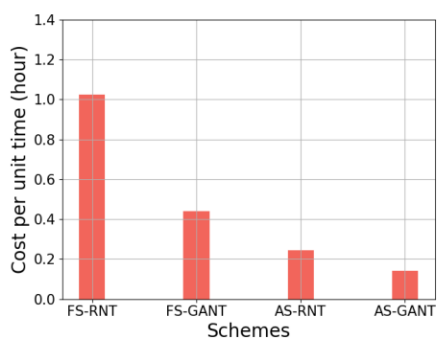

(c) Average defense cost per time unit ($\overline{C_D}$)

**Fig. 10.5: Performance comparison of the four MTD schemes with low-intelligent attackers where two decoy nodes are deployed for each VLAN.**

Fig. 10.5 shows the performance of the four schemes under this scenario (i.e., two decoy nodes for each VLAN with low-intelligent attackers) in terms of the mean values of three metrics based on the results collected from 100 simulation runs. As expected, Fig. 10.5 (a) shows the results similar to what we observed in Fig. 10.3 (a) and Fig. 4 (a). The random shuffling-based schemes, both FS-RNT and GA-RNT, have a similar number of attack paths towards decoy targets which is roughly 52; with the GA-based shuffling, $\overline{\overline{N_{DT}^{AP}}}$ is much higher, showing 85 with FS-GANT and 88 with AS-GANT. As more decoys are deployed in the network, all four schemes have higher $\overline{\overline{N_{DT}^{AP}}}$ compared with the schemes when only one decoy is deployed in each VLAN.

Fig. 10.5 (b) also follows the similar trends to the results shown in Fig. 10.3 (b) and Fig. 10.4 (b). With FS, both schemes have higher $\overline{MTTSF}$ compared with AS schemes. Besides, FS-GANT scheme has higher $\overline{MTTSF}$ (i.e., 1,757 hours) than FS-RNT scheme (i.e., 1,423 hours) while AS-GANT scheme also has slightly higher $\overline{MTTSF}$ than AS-RNT scheme, demonstrating 809 and 831 hours, respectively. Due to more decoys deployed in the network, all four schemes have higher $\overline{MTTSF}$ than a single decoy being deployed in each VLAN.

Fig. 10.5 (c) shows a slightly different trend compared to the results in Fig. 10.3 (c) and Fig. 10.4 (c). As expected, FS-RNT incurs the highest cost (i.e., 1.03) among all while FS-GANT incurs the second highest cost (i.e., 0.44) due to more frequent executions of network shuffling than adaptive shuffling. AS-GANT has the lowest cost (i.e., 0.14) among all. As more edges are available to be changed due to the increased number of decoy nodes, both FS-RNT and FS-GANT schemes incur higher cost than AS counterparts. Due to more connections that could be shuffled between decoys and real nodes, all four schemes under two decoys in each VLAN incur higher cost than the corresponding schemes under one decoy deployed in each VLAN.

Overall, GA-based schemes can provide a higher security level in terms of $N_{DN}^{AP}$. Adaptive shuffling-based schemes incur lower cost while fixed shuffling-based schemes incur higher $\overline{MTTSF}$. We can see there is a balance between MTTSF and defense cost per time unit. $\overline{MTTSF}$ of FS-GANT is roughly 2.1 times higher than $\overline{MTTSF}$ of AS-GANT while $\widehat{\overline{C_D}}$ of FS-GANT is 3.1 times higher than that one of AS-GANT. Compared with the results in Section 10.5.5.1, the network with an increasing number of decoys has higher security level but requires more cost due to a greater number of potential connections for shuffling.

## 10.6 Conclusions and Future Work

In this work, we have proposed an integrated proactive defense mechanism for an SDN-based IoT environment based on cyberdeception and MTD to improve security and mitigate the impact of potential attacks. More specifically, we have developed a network topology-based shuffling (NTS) in a network deployed with decoy nodes. We adopted a genetic algorithm (GA) and devised fitness functions to consider three system objectives: maximizing security by maximizing a number of attack paths towards decoy targets and mean time to security failure while minimizing defense cost derived from network topology shuffling operations. We also considered adaptive shuffling by introducing a metric to detect the system security vulnerability level based on the number of detected, compromised nodes and the number of critical nodes being compromised. For the assessment of our proposed NTS-MTD, we employed a graphical security model, namely HARM, based on the combination of attack graphs and attack trees. Via our extensive simulation study, we devised the four schemes considering either adaptive shuffling or fixed shuffling to determine when to trigger an MTD operation while answering how to trigger an MTD operation by using a random shuffling or GA-based intelligent shuffling. We compared the performance of the four schemes under three different scenarios by varying the number of decoys and the attacker's intelligence levels in detecting the decoys. Finally, we observed the outperformance of GA-based shuffling regarding the number of attack paths towards decoy targets and the balance between MTTSF and defense cost for fixed and adaptive GA-based shuffling schemes under the extensive performance analysis [29].

In our future work, we plan to explore: (1) how to apply our proposed scheme in large-scale IoT networks with a variety of decoy nodes by showing high scalability; (2) carrying out sensitivity analysis by varying the values of other key design parameters (e.g., weights to consider each system objective, a more number of decoy nodes deployed in the network, a more number of attackers, and/or system security vulnerability thresholds); (3) investigating the effect of deception and/or MTD on service availability, such as delay introduced by the deployment of decoy nodes and/or network topology shuffling; and (4) identifying an optimal setting of adaptive network topology shuffling algorithms in terms of thresholds in detecting system vulnerabilities.

**References**

[1]  H. Abie and I. Balasingham, "Risk-based adaptive security for smart IoT in eHealth," in *Proceedings of the 7th International Conference on Body Area Networks*, 2012, pp. 269-275: ICST (Institute for Computer Sciences, Social-Informatics and Telecommunications Engineering).

[2]  M. Anirudh, S. A. Thileeban, and D. J. Nallathambi, "Use of honeypots for mitigating DoS attacks targeted on IoT networks," in *Computer, Communication and Signal Processing (ICCCSP), 2017 International Conference on*, 2017, pp. 1-4: IEEE.

[3] C. J. Bernardos *et al.*, "An architecture for software defined wireless networking," *IEEE wireless communications,* vol. 21, no. 3, pp. 52-61, 2014.

[4] V. Casola, A. De Benedictis, and M. Albanese, "A moving target defense approach for protecting resource-constrained distributed devices," in *Information Reuse and Integration (IRI), 2013 IEEE 14th International Conference on*, 2013, pp. 22-29: IEEE.

[5] J.-H. Cho and N. Ben-Asher, "Cyber defense in breadth: Modeling and analysis of integrated defense systems," *The Journal of Defense Modeling and Simulation,* vol. 15, no. 2, pp. 147-160, 2018.

[6] J. Cho, Y. Wang, I. Chen, K. S. Chan, and A. Swami, "A Survey on Modeling and Optimizing Multi-Objective Systems," *IEEE Communications Surveys Tutorials,* vol. 19, no. 3, pp. 1867-1901, 2017.

[7] B. T. De Oliveira, L. B. Gabriel, and C. B. Margi, "TinySDN: Enabling multiple controllers for software-defined wireless sensor networks," *IEEE Latin America Transactions,* vol. 13, no. 11, pp. 3690-3696, 2015.

[8] S. Dowling, M. Schukat, and H. Melvin, "A ZigBee honeypot to assess IoT cyberattack behaviour," in *Signals and Systems Conference (ISSC), 2017 28th Irish*, 2017, pp. 1-6: IEEE.

[9] I. Farris, T. Taleb, Y. Khettab, and J. S. Song, "A survey on emerging SDN and NFV security mechanisms for IoT systems," *IEEE Communications Surveys & Tutorials,* 2018.

[10] L. Galluccio, S. Milardo, G. Morabito, and S. Palazzo, "SDN-WISE: Design, prototyping and experimentation of a stateful SDN solution for WIreless SEnsor networks," in *Computer Communications (INFOCOM), 2015 IEEE Conference on*, 2015, pp. 513-521: IEEE.

[11] F. C. Gärtner, "Byzantine failures and security: Arbitrary is not (always) random," 2003.

[12] M. Ge, J. B. Hong, W. Guttmann, and D. S. Kim, "A framework for automating security analysis of the internet of things," *Journal of Network and Computer Applications,* vol. 83, pp. 12-27, 2017.

[13] M. Ge, J. B. Hong, S. E. Yusuf, and D. S. Kim, "Proactive defense mechanisms for the software-defined Internet of Things with non-patchable vulnerabilities," *Future Generation Computer Systems,* vol. 78, pp. 568-582, 2018.

[14] J. B. Hong and D. S. Kim, "Assessing the effectiveness of moving target defenses using security models," *IEEE Transactions on Dependable and Secure Computing,* no. 1, pp. 1-1, 2016.

[15] F. Jia, J. B. Hong, and D. S. Kim, "Towards automated generation and visualization of hierarchical attack representation models," in *Computer and Information Technology; Ubiquitous Computing and Communications; Dependable, Autonomic and Secure Computing; Pervasive Intelligence and Computing (CIT/IUCC/DASC/PICOM), 2015 IEEE International Conference on*, 2015, pp. 1689-1696: IEEE.

[16] Q. D. La, T. Q. Quek, J. Lee, S. Jin, and H. Zhu, "Deceptive attack and defense game in honeypot-enabled networks for the internet of things," *IEEE Internet of Things Journal,* vol. 3, no. 6, pp. 1025-1035, 2016.

[17] T. Lei, Z. Lu, X. Wen, X. Zhao, and L. Wang, "SWAN: An SDN based campus WLAN framework," in *Wireless Communications, Vehicular Technology, Information Theory and Aerospace & Electronic Systems (VITAE), 2014 4th International Conference on*, 2014, pp. 1-5: IEEE.

[18] J. Liu, Y. Li, M. Chen, W. Dong, and D. Jin, "Software-defined internet of things for smart urban sensing," *IEEE communications magazine,* vol. 53, no. 9, pp. 55-63, 2015.

[19] K. Mahmood and D. M. Shila, "Moving target defense for Internet of Things using context aware code partitioning and code diversification," in *2016 IEEE 3rd World Forum on Internet of Things (WF-IoT)*, 2016, pp. 329-330: IEEE.

[20] T. Miyazaki *et al.*, "A software defined wireless sensor network," in *Computing, Networking and Communications (ICNC), 2014 International Conference on*, 2014, pp. 847-852: IEEE.

[21] NIST, "National Vulnerability Database (NVD). ," 2005, Available: https://nvd.nist.gov/.

[22]   L. Pingree, "Emerging technology analysis: Deception techniques and technologies create security technology business opportunities," *Gartner, Inc,* 2015.

[23]   R. Roman, J. Zhou, and J. Lopez, "On the features and challenges of security and privacy in distributed internet of things," *Computer Networks,* vol. 57, no. 10, pp. 2266-2279, 2013.

[24]   A. Rullo, E. Serra, E. Bertino, and J. Lobo, "Shortfall-Based Optimal Security Provisioning for Internet of Things," in *2017 IEEE 37th International Conference on Distributed Computing Systems (ICDCS)*, 2017, pp. 2585-2586: IEEE.

[25]   V. Saini, Q. Duan, and V. J. J. o. C. S. i. C. Paruchuri, "Threat modeling using attack trees," vol. 23, no. 4, pp. 124-131, 2008.

[26]   R. M. Savola, H. Abie, and M. Sihvonen, "Towards metrics-driven adaptive security management in e-health IoT applications," in *Proceedings of the 7th International Conference on Body Area Networks*, 2012, pp. 276-281: ICST (Institute for Computer Sciences, Social-Informatics and Telecommunications Engineering).

[27]   M. Sherburne, R. Marchany, and J. Tront, "Implementing moving target ipv6 defense to secure 6lowpan in the internet of things and smart grid," in *Proceedings of the 9th Annual Cyber and Information Security Research Conference*, 2014, pp. 37-40: ACM.

[28]   O. Sheyner, J. Haines, S. Jha, R. Lippmann, and J. M. Wing, "Automated generation and analysis of attack graphs," in *null*, 2002, p. 273: IEEE.

[29]   K. S. Trivedi. (2011). *Software packages: SHARP & SPNP*. Available: http://trivedi.pratt.duke.edu/software-packages.

[30]   K. Zeitz, M. Cantrell, R. Marchany, and J. Tront, "Designing a Micro-Moving Target IPv6 Defense for the Internet of Things," in *Internet-of-Things Design and Implementation (IoTDI), 2017 IEEE/ACM Second International Conference on*, 2017, pp. 179-184: IEEE.

[31]   OpenFlow Switch Specification (Version 1.3.0). Technical Report, ONF, 2012.